\documentclass[11pt,twoside]{article}   
\usepackage{epsf}
\date{}
\title{Energy Extraction from a Kerr Black Hole - An Ultimate Power Source in the Universe?}
\author{Amir Levinson\\ School of Physics and Astronomy\\ Tel Aviv University\\Tel Aviv 69978, Israel\\
Levinson@wise.tau.ac.il}
\begin{document}
\maketitle 
\begin{abstract}
It is widely believed that some classes of high-energy transients may be 
powered by the rotational 
energy of a rapidly spinning  black hole.   The energy extraction mechanism 
commonly discussed involves macroscopic magnetic fields that are produced by 
currents flowing in a disk or torus surrounding the black hole.  
The discovery of relativistic jets in radio loud AGNs nearly half a 
century ago was the main motivation for the analysis presented in the seminal paper 
by Blandford and Znajeck, where a global model for the magnetosphere of a Kerr 
black hole in the force-free limit was first constructed and employed 
to demonstrate that, under reasonable astrophysical conditions, energy can be 
extracted quite efficiently in the form of a Poynting flux.  
The recent discoveries of relativistic motions in Galactic microquasars, and 
the indications that GRBs eject ultra-relativistic, collimated outflows,  have
lent support to the hypothesis that this mechanism may be universal. 

In spite of large efforts that have led to impressive progress in our understanding
of the physics of magnetized flows in Kerr geometry, several important issues remain 
unresolved: the nature of the load in the Blandford-Znajeck model,
the causality and stability of a force free magnetosphere, global current closure,
and the role of boundary conditions on the channels through which the extracted
energy is released.   This chapter provides an account of recent work that addresses
some of those open questions.
After a brief introduction, an overview of MHD in Kerr geometry will be given. 
The basic properties of a static black hole magnetosphere will be 
discussed, with particular emphasis on the conditions under which efficient 
energy extraction may occur.  The trans-field equation that determines the geometry 
of magnetic flux surfaces, and the role of its critical surfaces will 
be summarized.
The force-free limit will be considered, with an attempt to clarify several issues 
concerning the long standing causality problem.   This will be followed by a 
discussion on global magnetospheric structures and related emission channels.
A particular model, developed recently for the interaction of a uniformly magnetized torus with 
a rapidly spinning black hole under extreme conditions, those anticipated in magnetars and GRBs,
will be considered in some greater detail.  The magnetic field
configuration invoked in this model is vastly different from the standard one 
adopted in previous work, and leads to interesting astrophysical consequences that 
will be described in detail.  In particular, the model predicts that the major fraction 
of the black hole spin
energy should be released in the form of gravitational waves, with estimated 
detection rate of up to a few per year with the LIGO and VIRGO detectors.
Non-stationary effects will be considered in the last part of this chapter.
The requirements for local adjustments of a magnetic flux tube to temporal changes will
be carefully examined, and it will be argued that changes in the angular velocity
of magnetic lines induced by small disturbances, involve effects beyond
the lowest order geometric optical approximation.
The global evolution of a force-free magnetosphere will be explored next, with an
attempt to elucidate the critical role played by the frame-dragging dynamo.
Recent analytic results will be presented and contrasted
with full GRMHD simulations. 
\end{abstract}
\section{Introduction}
It is widely believed that some classes of high-energy transients, notably AGNs, microquasars, and 
gamma-ray bursts (GRBs), may be powered by the rotational energy of 
a rapidly spinning black hole. The energy extraction mechanism commonly discussed involves macroscopic magnetic
fields that are produced by currents flowing in a disk or torus surrounding the black hole.  
A version of this idea was proposed originally by Ruffini \& Wilson (1975).  Blandford \& Znajek 
(1977; henceforth BZ) were the 
first to construct a global model for the magnetosphere of a Kerr black hole in the force-free limit, and 
to demonstrate that energy can be extracted under certain conditions in the form of a Poynting flux.
They applied their model to radio loud AGNs, and proposed that extragalactic jets are the consequence of
this energy extraction mechanism (see review by Begelman et al. 1984).  The dissipation of the Poynting 
flux is not addressed within the framework of the BZ model; it is merely assumed to occur in some 
non-force-free region where the magnetic field is sufficiently weak.  The detection of prodigious
gamma-ray emission from blazars by EGRET on CGRO (Thompson et al. 1993) 
indicates that a considerable fraction of the bulk energy is dissipated already on rather small scales. 
Conversion of magnetic energy to kinetic energy is a feature that appears to be common to many
classes of relativistic astrophysical systems.  The mechanism responsible for this process is poorly
understood at present, although some ideas have been discussed (Romanova \& Lovelace 1992, 1997; 
Levinson 1996, 1998; Thomson 1997; Levinson \& van Putten 1997; Kirk \& Lyubarsky 2001; Lyubarsky 2003).  
An important diagnostic of the BZ model is the 
composition of the collimated outflows, which is expected to be dominated by an 
electron-positron plasma.  Unfortunately, the content of extragalactic as well as Galactic
jets is yet an open issue (e.g., Mannheim 1993; Dermer \& Shlickheiser 1993; Sikora et al. 1994; Blandford \&
Levinson 1995; Celotti 1997).  Some constraints can be imposed from the 
observations (Sikora et al. 1997; Celotti 1997), but the analysis is inconclusive.
This issue may be eventually resolved by the upcoming km$^3$ neutrino detectors.
Various aspects of the BZ model have 
been subsequently examined in more detail by different authors (e.g., Bekenstein \& Oron 1978; 
Macdonald \& Thorne 1982; Phinney 1983; Thorne et al. 1986; Punsly \& Coroniti, 1990; 
Ghosh \& Abramovitz 1997), but many issues remained to be resolved. 

The BZ model has been applied in recent years also to stellar systems, in particular
GRBs (e.g., Levinson \& Eichler 1993; Meszaros \& Rees 1997; 
Van Putten 2000, 2001; Lee, et al. 2000; Brown, et al. 2000), and microquasars (Levinson \& Blandford 1996).
The formation of systems consisting of a rapidly rotating, stellar mass black hole
surrounded by a magnetized torus, is thought to be the outcome of catastrophic events such as black hole-
neutron star and neutron star-neutron star coalescence (e.g., Eichler et al. 1989; Paczy\`nski 1991) or
core collapse of massive stars (Woosley 1993).  The putative existence of such black hole-torus 
remnants is by itself a strong motivation to consider the physics and observational consequences of 
the interaction of the black hole and the torus.  It is quite likely that upcoming gamma-ray (e.g., MAGIC, GLAST), 
km$^3$ neutrino (e.g., Ice-Cube, NEMO), and
gravitational wave (LIGO, VIRGO) detectors will detect new signatures in currently identified
classes of high-energy transients, or may even discover new classes of astrophysical objects, that may be the 
product of the underlying black hole-torus systems.  A particular example is emission of gravitational
waves by a torus in suspended accretion state (Van Putten 2001, 2002; Van Putten \& Levinson 2002, 2003). 
A key feature in this model that distinguishes it from the original BZ scenario, 
is a strong coupling between the black hole and the torus which is mediated by angular momentum transfer 
along magnetic field lines connecting the horizon and the inner face of the torus.
The emission of the gravitational waves by the torus, as well as intense baryon rich winds is a consequence 
of this coupling.  This example suggests that remnants of collapsed,
massive stars may be prodigious sources of gravitational waves.  The proposed association of these objects with 
presently known classes of GRBs (Van Putten 2001, Van Putten \& Levinson 2001), is motivated in this model both 
phenomenologically and from theoretical considerations that imply that a small fraction of the black hole rotational 
energy should be expelled along the axis in the form of ultra-relativistic, baryon poor jets.  However, the 
properties of the emission from those baryon poor jets are likely to depend also on environmental
conditions and additional microphysics, and it could well be that the association with identified classes of GRBs 
applies only to a sub-class of these potential gravitational wave sources.  Moreover, the prompt 
GRB emission is presumably strongly beamed, while the gravitational wave emission is roughly isotropic, so that most of those 
putative gravitational wave sources are not expected to be coincide with an observed GRB anyhow.  On the other
hand, delayed, isotropic radio emission is anticipated from the interaction of the torus winds with the 
ambient medium, and/or from the GRB afterglow.  Systematic searches for radio transients may eventually 
reveal theses remnants (Levinson et al. 2002).  A search for associations of radio transients with LIGO signals may 
provide valuable information and may aid detecting these sources.  

The total amount of energy released in the BZ process depends mainly on the mass, $M$, and angular 
momentum, $a$, of the black hole, and the corresponding rate depends 
on the strength of the poloidal magnetic field threading the horizon, $B$.  To be more precise,
the Blandford-Znajek power can be expressed as (see section \ref{sec:FF}):
\begin{equation}
P_{BZ}=\epsilon \left({a\over M}\right)^2B^2r_H^2c,
\label{P_BZ}
\end{equation}
where $r_H=2GM/c^2$ is the Schwartchild radius, and $\epsilon$ is an efficiency factor 
that depends on the geometry of the magnetic field, as will be shown below.  
Typically, $\epsilon$ lies in the range $10^{-2}-10^{-1}$. 
For a supermassive black hole of mass $M\sim10^9$ solar masses, as expected in the 
powerful blazars, magnetic field of the order of $10^4$ G is required to explain the
energetics of the associated gamma-ray jets.   The magnetic field may be supported by a surrounding
disk and extend close to the event horizon (e.g., Bisnovatyi-Kogan \& Ruzmaikin 1976).  
In the case of GRBs, that harbor stellar mass black holes,
super-critical magnetic fields, of order $10^{15}$ G are needed in order to account for the 
characteristic luminosities measured.  Such fields are much stronger than those inferred in 
typical radio pulsars.  Which process can give rise to 
such amplification of the magnetic field is yet an open issue, although some models have been proposed 
(e.g., Thomson \& Duncan 1993).
Recent observational efforts provide indications of supercritical magnetic fields in magnetars 
(Duncan \& Thompson 1992; Kouveliotou et al. 2003), 
suggesting that such high magnetic fields may also be present in other extreme systems.

\section{ IDEAL MHD IN KERR GEOMETRY}
\subsection{Basic equations and integrals of motion}
We express the Kerr metric (Kerr 1963) in Boyer-Lindquest coordinates with the following notation:
\begin{equation}
ds^2= -\alpha^2dt^2 + \tilde{\omega}^2(d\phi+\beta dt)^2 + \frac{\rho^2}{\Delta}dr^2 + 
\rho^2 d\theta^2,
\label{metric}
\end{equation}
where $\alpha=\rho\sqrt{\Delta}/\Sigma$ is the lapse function, 
$\tilde{\omega}^2=(\Sigma^2/\rho^2)\sin^2\theta$, and
$-\beta=2aMr/\Sigma^2$ is the angular velocity of a ZAMO with respect to a distant 
observer , with $\Delta=r^2+a^2-2Mr$, $\rho^2=r^2+a^2\cos^2\theta$, and 
$\Sigma^2=(r^2+a^2)^2-a^2\Delta\sin^2\theta$.  The parameters $M$ and $a$ are the mass
and angular momentum per unit mass of the black hole.

We denote by $n$, $p$, $\rho$, $h=(\rho+p)/n$, respectively, the proper particle 
density, pressure, energy density, and specific enthalpy of the MHD flow.  
The stress-energy tensor then takes the form:
\begin{equation}
T^{\alpha\beta} = hnu^{\alpha}u^{\beta} + pg^{\alpha\beta}+\frac{1}{4\pi}
(F^{\alpha\sigma}F^{\beta}_{\sigma}-\frac{1}{4}g^{\alpha\beta}F^2),
\label{Tmunu}
\end{equation}
where $u^{\alpha}$ is the four-velocity, and $F_{\mu\nu}=\partial _\mu A_\nu
-\partial _\nu A_\mu$ is the electromagnetic 
tensor.  The dynamics of the MHD system is then governed by the following set of equations: 
Maxwell's equations,
\begin{eqnarray}
F^{\beta\alpha}_{;\alpha}=\frac{1}{\sqrt{-g}}(\sqrt{-g}F^{\beta\alpha})_{,\alpha}
=4\pi j^{\beta},\label{F=j}\\
F_{\alpha\beta,\gamma}+F_{\beta\gamma,\alpha}+F_{\gamma\alpha,\beta}=0,\label{F=0}
\end{eqnarray}
the continuity equation
\begin{equation}
(nu^{\alpha})_{;\alpha}=\frac{1}{\sqrt{-g}}(\sqrt{-g}nu^{\alpha})_{,\alpha}=0,
\label{continuity}
\end{equation}
and the energy and momentum equations
\begin{equation}
T^{\mu\nu}_{\ \ ;\nu}=\frac{1}{\sqrt{-g}}(\sqrt{-g}T^{\mu\nu})_{,\nu}+
\Gamma^{\mu}_{\alpha\beta}T^{\alpha\beta} = 0.
\label{Tmn}
\end{equation}
Here $\Gamma^{\mu}_{\alpha\beta}$ denotes the associated Cristofel symbol 

In regions where the MHD flow is ideal (i.e., infinite conductivity), Ohms law
yields the additional constraint,
\begin{equation}
u^{\alpha}F_{\alpha\beta}=0.
\label{ideal}
\end{equation}
The stationary axisymmetric flow considered here is characterized by two Killing vectors:
$\xi^{\mu}=\partial_t$ and $\chi^{\mu}=\partial_{\phi}$.  By contracting these Killing 
vectors with the stress-energy tensor we can construct the energy and angular momentum 
currents, which in Boyer-Lindquest coordinates read:
\begin{eqnarray}
{\cal E}^{r}=T^{r}_{\beta}\xi^{\beta}=-T^r_t=-hnU^rU_t-\frac{1}{4\pi}F^{r\theta}F_{t\theta},\label{Er}\\
{\cal E}^{\theta}=T^{\theta}_{\beta}\xi^{\beta}=-T^{\theta}_t=-hnU^{\theta}U_t+\frac{1}{4\pi}
F^{r\theta}F_{t r},\label{Etheta}\\
{\cal L}^{r}=T^{r}_{\beta}\chi^{\beta}=T^r_{\phi}=hnU^rU_{\phi}+\frac{1}{4\pi}
F^{r\theta}F_{\phi\theta},\label{Lr}\\
{\cal L}^{\theta}=T^{\theta}_{\beta}\chi^{\beta}=T^{\theta}_{\phi}=hnU^{\theta}U_{\phi}-
\frac{1}{4\pi}F^{r\theta}F_{\phi r}.
\label{Ltheta}
\end{eqnarray}
These currents are conserved, viz.,
\begin{equation}
{\cal L}^{\alpha}_{;\alpha}={\cal E}^{\alpha}_{;\alpha}=0.
\label{currents}
\end{equation}

The relation 
\begin{equation}
0=F_{\phi\theta}F_{\phi r}+F_{r\phi}F_{\phi\theta}=-(F_{\phi\theta}
\partial_r + F_{r\phi}\partial_{\theta})A_{\phi}\equiv D_{\psi}A_{\phi},
\label{DA_phi}
\end{equation}
where the operator $D_{\psi}=F_{\theta\phi}\partial_r + F_{\phi r}\partial_{\theta}$ 
denotes derivative along magnetic flux surfaces, implies that $A_{\phi}$ is conserved 
on magnetic flux surfaces and is, therefore, a viable stream function.
We can use the dual electromagnetic tensor, denoted here by ${\cal F}^{\mu\nu}$, to
express the invariant ${\bf E}\cdot{\bf B}$ in the form,
\begin{equation}
{\bf E}\cdot{\bf B}=- (\sqrt{-g}/4){\cal F}^{\mu\nu}F_{\mu\nu}= D_{\psi}A_t. 
\label{EB}
\end{equation}
It can be readily shown that the ideal MHD condition, given by eq. (\ref{ideal}), implies
that ${\bf E}\cdot{\bf B}=0$, meaning that the electric potential is conserved
along magnetic surfaces.  We note that in cases where a gauge $A_t=-\Omega A_{\phi}$
can be found, where $\Omega(r,\theta)$ is some function of the coordinates, the 
change in the electric potential along flux surfaces is given by 
$D_{\psi}A_t=-A_{\phi}D_{\psi}\Omega$, where eq. (\ref{DA_phi}) has been used.
Thus, in regions where the MHD flow is ideal $\Omega$ is conserved along magnetic surfaces.
As discussed below the function $\Omega$ is approximately the angular velocity of the magnetic
flux tube at small angles.  On the other hand, in regions where the ideal MHD condition is violated, e.g.,
the sparking gap (see below), the change in angular velocity across this region is
related to the electric potential drop through: $\Delta A_t=-A_{\phi}\Delta \Omega$ (note
that in steady-state $A_\phi$ is conserved even in the non-ideal case).  The relevancy of this
result to the double transonic flow is discussed below.

Equations (\ref{F=0}), (\ref{continuity}), (\ref{ideal}), and (\ref{Er})-(\ref{currents}) admit 
four additional quantities that are conserved along magnetic flux surfaces 
(e.g., Bekenstein \& Oron 1978):  The particle flux per unit magnetic flux,
\begin{equation}
\eta(\Psi)=\frac{\sqrt{-g}nu^r}{F_{\theta\phi}}=\frac{\sqrt{-g}nu^{\theta}}{F_{\phi r}};
\label{eta}
\end{equation}
the angular velocity of magnetic field lines,
\begin{equation}
\Omega_F(\Psi)=-{\eta\over \sqrt{-g}nu^t}F_{r\theta}+v^\phi;
\label{Omega}
\end{equation}
and the total energy and angular momentum per particle carried by the MHD flow,
\begin{eqnarray}
\label{E}
E(\Psi)=-h u_t -\frac{\sqrt{-g}}{4\pi\eta}\Omega_F F^{r\theta},\\
\label{L}
L(\Psi)=h u_{\phi}-\frac{\sqrt{-g}}{4\pi\eta} F^{r\theta}.
\end{eqnarray}
Here $v^{\phi}=u^{\phi}/u^{t}$, and $v^r=u^r/u^t$ are the corresponding components 
of the 3-velocity.   The fifth integral of motion is the entropy $s=s(\Psi)$.
Equations (\ref{eta})- (\ref{L}) can be solved for $u_t$, $u_\phi$
and $F^{r\theta}$.  The solution can then be used to express  $F^{r}_{\theta}$, 
$u^t = g^{tt}u_t+g^{t\phi}u_\phi$, and
$u^{\phi}= g^{\phi t}u_t+g^{\phi\phi}u_\phi$ as:
\begin{eqnarray}
F^{r}_{\theta}= -\frac{4\pi\eta}{\sin\theta}\frac{\alpha^2L-\tilde{\omega}^2
(\Omega_F+\beta)(E+\beta L)}{\alpha^2-\tilde{\omega}^2(\Omega_F+\beta)^2 - M^2},
\label{F^r_tet}\\
hu^t =\frac{1}{\alpha^2}\frac{\alpha^2(E-\Omega_F L)-M^2(E+\beta  L)}
{\alpha^2-\tilde{\omega}^2(\Omega_F+\beta)^2 - M^2},\label{u^t}\\
hu^{\phi}= \frac{\alpha^2\tilde{\omega}^2\Omega_F
(E-\Omega_F L)+\tilde{\omega}^2\beta M^2(E+\beta L)
-\alpha^2M^2 L}{\alpha^2\tilde{\omega}^2[\alpha^2-\tilde{\omega}^2(\Omega_F+\beta)^2- M^2]},
\label{u^phi}
\end{eqnarray}
where $M^2=(4\pi h\eta^2)/n$.  The meaning of the latter becomes clear when written in terms of
ZAMO 4-velocities.  By employing eq. (\ref{eta}) one obtains
$M^2=\alpha^2 u_p^2/u_A^2$, where $u_p=(u^ru_r+u^\theta u_\theta)^{1/2}$ 
and $u_A=(B_p^2/4\pi h n)^{1/2}$, with $B_p^2=(\Sigma\sin\theta)^{-2}
(F_{\phi\theta}^2+\Delta F_{\phi r}^2)$, are, respectively,  the poloidal velocity and Alfv\'en 
velocity, as measured by a ZAMO (see e.g., Macdonald \& Thorne 1982, Beskin 1997).
Thus, $M$ is up to a factor $\alpha$ the Alfven Mach number.
Finally, using the normalization condition, $u^{\mu}u_{\mu}=-1$ and the definition of 
the Mach number $M$, we obtain the Bernoulli equation
\begin{equation}
u_p^2+1=\frac{M^4B_p^2}{16\pi^2\eta^2\alpha^2h^2}+1=(\alpha u^t)^2-\tilde{\omega}^2(\beta u^t+u^\phi)^2.
\label{Berno}
\end{equation}
Note that since the enthalpy $h$ is a function of the thermodynamic variables $n$ and $s$ only,
the density and enthalpy can be expressed in terms of $s$ and $M$, viz., $n=n(s,M)$, $h=h(s,M)$.
Thus, equations (\ref{F^r_tet})-(\ref{Berno}) determine essentially all the flow characteristics
in terms of the five integrals of motion once the poloidal magnetic field is known. 

\subsection{Asymptotic behavior of the MHD flow}

Consider first the behavior of the solution near the horizon.  There $\alpha\rightarrow 0$,
and for physical solutions for which the Mach number is finite on the horizon, $M(r_H)\ne0$,
eqs. (\ref{u^t}) and (\ref{u^phi}) yield $u^\phi\rightarrow\Omega_H u^t$ as $\alpha\rightarrow0$,
where $\Omega_H=-\beta_H$ is the angular velocity of the black hole.  Consequently, the rotation of the 
plasma on the horizon is synchronous with the black hole, as one might expect. Substituting the
latter result into Bernoulli 
equation (\ref{Berno}), we obtain $u_p\rightarrow \alpha u^t$ on the horizon. 
The poloidal velocity is radial on the horizon, implying $u_p^2 \rightarrow  g_{rr}u^ru^r$, 
hence
\begin{equation}
v^r =u^r/u^t \rightarrow \alpha/\sqrt{g_{rr}} = -\Delta/(r^2+a^2).
\end{equation}
This shows that the plasma moves along geodesics of a freely falling observer as it approaches
the horizon, meaning that near the horizon the dynamics is governed by gravity alone.
Substituting $v^\phi$ and $v^r$ obtained above into eqs. (\ref{eta}) and (\ref{Omega}), 
yields
\begin{equation}
\frac{F_{r\theta}}{F_{\phi\theta}}= - \frac{r^2+a^2}{\Delta}(\Omega_F-\Omega_H). 
\label{ratio1}
\end{equation}
Eq. (\ref{ratio1}) gives the frozen-in condition derived originally by Znajek (1977) 
and used by BZ as a boundary condition in their force-free analysis.   

Next, consider the behavior of a MHD outflow far from the black hole.   The outflow parameters
are given, to a good approximation, by eqs (\ref{F^r_tet})-(\ref{Berno}) with 
$\alpha^2=1$, $\beta=0$.  At the 
Alfv\'en surface the Mach number is $M_A^2=1-\tilde{\omega}_A^2\Omega_F^2$, and the 
requirement that $F^r_\theta$ $u^t$ and $u^\phi$ remain finite there yields the well known 
relation  $L/E=\tilde{\omega}_A^2\Omega_F$ (e.g., Weber \& Davis 1967; Camenzind 1986).
By employing eqs. (\ref{eta}) and (\ref{Omega}) we obtain,
\begin{equation}
\frac{F_{r\theta}}{F_{\phi\theta}}=-\frac{M^2(R^2\Omega E-L)}{R^2 v^r[(1-M^2)E-\Omega L]}=
\frac{(\tilde{\omega}^2-\tilde{\omega}_A^2)M^2\Omega_F}
{v^r\tilde{\omega}^2(M^2-M_A^2)}.
\end{equation}
Well above the Alfv\'en point, where $\tilde{\omega}>>\tilde{\omega}_A$ and $M>>M_A$
the last equation gives 
\begin{equation}
\frac{F_{r\theta}}{F_{\phi\theta}}\rightarrow\frac{\Omega_F}{v^r}.
\label{ratio2}
\end{equation}
The asymptotic poloidal current follows from the $r$ component of eq. (\ref{F=j}):
\begin{equation}
\frac{\partial}{\partial\theta}\left(\frac{\Delta\sin\theta}{\rho^2}
F_{r\theta}\right)=4\pi\sqrt{-g}j^r.
\label{Max-r}
\end{equation}
Integrating the latter equation we obtain the net electric current within a flux tube,
\begin{equation}
I=\int{2\pi\sqrt{-g}j^rd\theta}=\frac{\Delta \sin\theta}{2\rho^2}F_{r\theta}.
\label{I}
\end{equation}
Using eqs (\ref{ratio1}) and (\ref{ratio2}) we find that the net electric current flowing on the horizon is given by
\begin{equation}
I_H=\frac{\Sigma\sin\theta}{2\rho^2}(\Omega_H-\Omega_F)F_{\phi\theta},
\label{I_H}
\end{equation}
and the net electric current at infinity is given by, 
\begin{equation}
I_\infty=\frac{\Omega_F}{2v^r}\sin\theta F_{\phi\theta},
\label{I_inf}
\end{equation}

\subsection{Conditions for energy extraction by a magnetized inflow}
Detailed analysis of the inflow structure and the requirements
for extraction of the hole rotational energy is given in Takahashi et al. (1990)
and Hirotani et al. (1992).
For completeness, we give in this section a brief derivation of the conditions
under which energy extraction by a MHD inflow is possible.
From equations (\ref{Er})-(\ref{Ltheta}), (\ref{E}) and (\ref{L}) it is readily
seen that the energy and angular momentum fluxes can be expressed as:
${\cal E}^{a}=E nu^{a}$, and ${\cal L}^{a}=L nu^{a}$, with $a=r,\theta$.
Here $nu^{a}$ is the particle flux carried by the MHD flow.  Consequently, energy extraction
by the MHD inflow (for which $nu^\alpha$ is negative) then requires 
the specific energy to be negative, viz., $E(\psi)<0$.  This shows that the BZ mechanism is
indeed a manifestation of the Penrose process, as pointed out by Takahashi et al. (1990) 

Now, at the Alfv\'en surface the denominator of eqs (\ref{F^r_tet})-(\ref{u^phi}) vanishes,
and we obtain 
\begin{equation}
M_A^2=\alpha_A^2-\tilde{\omega}_A^2(\Omega_F+\beta_A)^2\ge 0,
\label{M_A}
\end{equation}
where the subscript $A$ refers to quantities on the Alfv\'en surface.  This in turn implies 
that the angular velocity of magnetic field lines must lie in the range 
\begin{equation}
-\beta_A-\alpha_A/\tilde{\omega}_A <\Omega_F< -\beta_A+\alpha_A/\tilde{\omega}_A.
\label{Omega-cond}
\end{equation}
In order for $u^t$ to be finite on the Alfv\'en surface, the relation
\begin{equation}
{L\over E}=\frac{\tilde{\omega}_A^2(\Omega_F+\beta_A)}
{\alpha_A^2-\tilde{\omega}_A^2\beta_A(\Omega_F+\beta_A)}
\label{E/L}
\end{equation}
must hold (see eq. [\ref{u^t}]).
At the injection surface of the inflow we take $M^2\sim0$.  Equation (\ref{Berno}) then yields
\begin{equation}
E-\Omega_FL=h_{inj}[\alpha_{inj}^2-\tilde{\omega}_{inj}^2(\Omega_F+\beta_{inj})^2]^{1/2}\ge 0.
\label{inj}
\end{equation}
Solving eqs (\ref{E/L}) and (\ref{inj}) we obtain
\begin{eqnarray}
E=\frac{\alpha_A^2-\tilde{\omega}_A^2\beta_A(\Omega_F+\beta_A)}
{\alpha_A^2-\tilde{\omega}_A^2(\Omega_F+\beta_A)^2}
h_{inj}[\alpha_{inj}^2-\tilde{\omega}_{inj}^2(\Omega_F+\beta_{inj})^2]^{1/2},
\label{Esol}\\
L=\frac{\tilde{\omega}_A^2(\Omega_F+\beta_A)}
{\alpha_A^2-\tilde{\omega}_A^2(\Omega_F+\beta_A)^2}
h_{inj}[\alpha_{inj}^2-\tilde{\omega}_{inj}^2(\Omega_F+\beta_{inj})^2]^{1/2}.
\label{Lsol}
\end{eqnarray}
Evidently, the specific energy will be negative provided the condition 
$\alpha_A^2-\tilde{\omega}_A^2\beta_A(\Omega_F+\beta_A)<0$ is satisfied 
on the Alfv\'en surface.  The latter condition combined with eq.
(\ref{Omega-cond}) implies, in turn, that the rotational energy of the hole 
can be extracted only if (i) the Alfv\'en surface of the inflow 
is located inside the ergosphere, and (ii) $0<\Omega_F<\Omega_H$.

\subsection{The trans-field equation}
\label{sec:GS}

Equation (\ref{eta}) can be used to express the poloidal components of the 4-velocity
in terms of the magnetic field components.  Substituting the latter into eq. (\ref{Tmunu}),
the poloidal components of the stress-energy tensor can be written as:
\begin{equation}
T^{ab}=\frac{M^2}{16\pi^2(\sqrt{-g})^2}(\partial_a \psi)(\partial_b\psi)-pg^{ab}+
\frac{1}{4\pi}
(F^{\alpha\sigma}F^{\beta}_{\sigma}+\frac{1}{4}g^{\alpha\beta}F^2),
\label{GS-T}
\end{equation}
where $\psi=2\pi A_\phi$ is the magnetic flux, and the indices $a,b$ run through
$r$ and $\theta$ only.  The poloidal magnetic and electric fields in eq. (\ref{GS-T}) 
are given in terms of the stream function $\psi$ as: 
$F_{\phi a}=-(1/2\pi)\partial_a \psi$, $F_{ta}=-\Omega_F F_{\phi a}$, and the toroidal
magnetic field $F_{r\theta}$ by eq. (\ref{F^r_tet}).  Upon substituting eq. (\ref{GS-T})
into eq. (\ref{Tmn}), one obtains a second order PDE for the stream function $\psi$, involving
the five integrals of motion, the Mach number, the proper density $n$ and the temperature.
The resultant trans-field equation is given explicitly in eq. (39) of Beskin (1997)\footnote{
The trans-field equation is written in Beskin 1997, who uses the 3+1 formalism, in 
terms of the operator ${\bf \nabla}=\hat{e}_r\sqrt{\Delta}\rho^{-1}\partial_r
+\hat{e}_\theta\rho^{-1}\partial_\theta$ }.  Combined with Bernoulli equation (\ref{Berno}) 
and an appropriate equation of state, that allow one
to express, though implicitly, the Mach number and the thermodynamic parameters in terms 
of the conserved quantities, this trans-field equation can be solved in principle 
to yield the stream function, $\psi=\psi(r,\theta)$, for a given set of boundary conditions.
The functional form of the invariants $\Omega(\psi)$, $L(\psi)$, $E(\psi)$, $\eta(\psi)$ 
and $s(\psi)$ that appear explicitly in the equation must be specified first through 
appropriate boundary conditions.

The trans-field equation has, in general, several singular surfaces.  Those singular surfaces
can be identified from differentiation of eq. (\ref{Berno}), whereby an expression for
the gradient of the Mach number can be obtained (Takahashi 1990; Beskin 1997).  
Of most interest are the Alfv\'en surface\footnote{It is noteworthy that although the Grad-Shafranov 
equation has a singularity on the Alfv\'en surface, the regularity of equations
(\ref{F^r_tet})-(\ref{u^phi}) is automatically satisfied.  It merely defines the location
of this surface.}
, on which the Mach number satisfies relation (\ref{M_A})
and the fast and slow magnetosonic surfaces on which the poloidal 4-velocity satisfies
\begin{equation}
u_p^2=K\pm\sqrt{K^2-4c_s^2u_A^2\left[1-(\Omega_F+\beta)^2\tilde{\omega}^2\alpha^{-2}\right]},
\end{equation}
with the $+$ ($-$) sign applies to the fast (slow) magnetosonic surface.  Here $2K=
[1-(\Omega_F+\beta)^2\tilde{\omega}^2\alpha^{-2}]u_A^2+(B_\phi^2/4\pi hn)+c_s^2$,
where $B_\phi=(\sqrt{\Delta}/\rho^2)F_{r\theta}$ is the ZAMO toridal magnetic field,
and $c_s$ is the sound 4-velocity, given by  $c_s^2=a_s^2/(1-a_s^2)$ with 
$a_s^2=(1/h)(\partial p/\partial n)_s$.  As the flow accelerates, it crosses
first the slow surface, then the Alfv\'en surface, next the light cylinder, 
and finally the fast surface.  
As shown in e.g., Beskin (1997), with the exception of the force-free case the trans-field
equation changes from elliptic to hyperbolic on the fast critical surface, meaning
that the flow beyond that surface (between the fast critical surface and the 
horizon in the inflow section) is not in causal contact with the region between the 
fast critical surface and the injection point.   In the force-free limit discussed next, 
the Alfv\'en surface coincides with the ligh cylinder, and the fast critical surface
with the horizon.  In this case the trans-field equation remains elliptic down to the horizon.
Now, the requirement that the MHD flow pass smoothly through the critical surface imposes 
certain relations between the conserved quantities of the MHD flow.  Since the trans-field
equation is second order the number of boundary conditions required is 2 + number of conserved
quantities - number of singular surfaces. 

There is ample literature on the properties of the Grad-Shafranov equation and its 
solutions in flat spacetime, mainly in the context of pulsar physics (see e.g., Mestel 1999,
and references therein).
Solutions of the trans-field equation in curved spacetime were obtained in various limits.
Examples are discussed in, e.g., (Blandford \& Znajek 1977; Fendt 1997; Beskin 1997; Ghosh 2000; 
Beskin \& Kuznetsova 2000;  Lee, at al. 2001; Uzdensky 2004a,b)

\subsection{The force-free limit}
\label{sec:FF}

In situations where the MHD flow is so highly magnetized that the 
inertia of the plasma can be ignored, the system approaches the 
force-free limit.  Formally this limit can be obtained by setting
$h\rightarrow0$ in the above equations \footnote{We note that in reality $h\ge mc^2$, where  
$m$ is the rest mass of a particle in the flow.  The limit $M^2<<1$
is due to the small particle density.  Thus, the limit of negligible inertia is better
characterized by the dimensionless Mach number.}.
This in turn implies $M^2\rightarrow0$ (meaning
essentially that the Alfv\'en velocity, as measured by a ZAMO, approaches 
the speed of light).
The Alfv\'en surface then approaches the light cylinder,
as can be readily seen from eq. (\ref{M_A}).
In the limit $h=0$ eq. (\ref{Tmn}) reduces to
\begin{equation}
F_{\mu\nu}j^{\nu}=0,
\label{FF}
\end{equation}
where eq. (\ref{F=j}) has been used.
The toroidal component of the last equation and the condition $F_{\phi t}=0$ 
give  $F_{\phi r}j^r+F_{\phi\theta}j^{\theta}=0$, meaning that the poloidal
current must flow along magnetic flux surfaces.  Consequently, the net electric 
current within a given flux surface, given by eq. (\ref{I}),
is conserved, as can be readily verified by using the $r$ and $\theta$ components
of eq. (\ref{F=j}) and the latter condition.   

The total energy and angular momentum are given by $E_{tot}=\eta E$ and $L_{tot}=\eta L$.   
By employing eqs. (\ref{E}) 
and (\ref{L}) we obtain $E_{tot}=\Omega_F L_{tot}$ (to be more precise, eq. [\ref{inj}] 
shows that this result is correct, in general, to order $h_{inj}\eta $).  
Taking the limit $M^2\rightarrow 0$ and $E\rightarrow\Omega_F L$ 
in eq. (\ref{F^r_tet}) and using eq. (\ref{I}) we also find $I=2\pi L_{tot}$.  Thus,
the force-free system has only two integrals of motion, $\Omega_F$ and $I$.
 
Combining eqs (\ref{Er}) and (\ref{ratio1}), one can obtain the energy flux on the horizon:
\begin{equation}
{\cal E}^r=\Omega_F(\Omega_H-\Omega_F)\frac{\Sigma}{\rho^4}(F_{\phi\theta})^2,
\label{calEr}
\end{equation}
which is a well known result, derived originally by BZ.  Maximum extraction 
efficiency occurs for $\Omega_F=\Omega_H/2$.  The net power transferred
outwards from the horizon is $P=2\pi\int{\sqrt{-g} {\cal E}^rd\theta}$, where
the integration is over the horizon surface.  By expressing $F_{\phi\theta}$ in terms 
of the ZAMO poloidal field, $B_r=-F_{\phi\theta}/\Sigma\sin\theta$, in eq. (\ref{calEr}),
and scaling the metric components, we arrive at eq. (\ref{P_BZ}), with the dimensionless
factor $\epsilon$ given explicitly as an integral involving the scaled components.

\subsection{The Double transonic flow}

\begin{figure}
\centerline{\epsfxsize=140mm\epsfbox{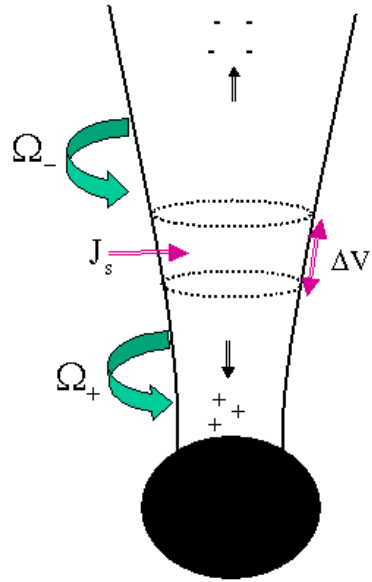}}
\caption{\small Schematic illustration of a double-transonic structure. Transonics 
inflow and outflow 
are expelled from the injection region where plasma is continuously generated. 
Magnetic flux surfaces are assumed to be continuous across the injection region. 
The angular velocities of magnetic field lines in the inflow and outflow sections 
are determined by the potential drop $\Delta V$ along magnetic surfaces in the injection 
region, and the cross-field electric current $J_s=I_H-I_{\infty}$ (see text).}
\end{figure}

In many applications of the BZ mechanism to astrophysical systems it is envisioned 
that the energy extracted near the horizon of the hole is ultimately transferred
to large radii in the form of a relativistic jet.  Since the poloidal velocity 
of the outflowing plasma is in the opposite direction to that of the negative 
energy inflow inside the ergosphere, the invariants $\eta(\psi)$, $E(\psi)$ and
$L(\psi)$ must have an opposite sign in the two regions (the energy and angular 
momentum fluxes can be continues nonetheless), implying that there must exist 
a region, that we shall term the injection region, where the ideal 
MHD condition is violated, and where fresh plasma is produced.  This region must 
be located somewhere between the inner and outer Alfv\'en surfaces (Takahashi et al. 1990).
We point out that 
this discontinuity in the flow does not appear implicitly in the force-free case,
where formally $\eta=0$.  As shown above, the force-free system has only two invariants,
$I$ and $\Omega$, both of which can be assumed continuous in the entire region between
the inner and outer fast magnetosonic surfaces.  The implicit assumption being made
is that there exists a source of electric charges, presumably located on the neutral 
surface defined below.  

For illustration let us consider the properties of the double transonic flow near the 
rotation axis of the black hole (figure 1).  We suppose that the MHD inflow and outflow expelled
from the injection region become ideal outside the injection region.  
The entire system is then characterized by 10 invariants, 5 for each flow.  We
henceforth refer to quantities in the inflow (outflow) by a subscript + (-).
Consider first the charge distribution near the axis.
The $t$ component of eq. (\ref{F=j}) reduces in the 
small angle approximation to,
\begin{equation}
\frac{\partial}{\partial\theta}\left[\frac{\sin\theta}{\alpha^2}
(\Omega+\beta)F_{\phi \theta}\right]=
4\pi\sqrt{-g}j^{t}.\label{Max t2}
\label{Max-t}
\end{equation}
For reasonable magnetic field configurations eq. (\ref{Max-t}) can be integrated to
yield the lowest order Goldriech-Julian charge density:
\begin{equation}
j^t=-\frac{(\Omega_F+\beta)B_r}{2\pi\alpha^2},
\label{j_GJ}
\end{equation}
where $B_r$ is the ZAMO radial magnetic field.  As seen, the charge density changes sign
across a neutral sheet on which the angular velocity of magnetic lines, as measured
by a ZAMO is zero, viz., $\Omega_F=-\beta$.   
Now, to lowest order eqs. (\ref{I_H}) and (\ref{I_inf}) give
\begin{eqnarray}
I_{H}(\psi)=-\frac{(\Omega_H-\Omega_{+})\psi}{2\pi},\\
I_{\infty}(\psi)=-\frac{\Omega_{-}\psi}{2\pi}.
\end{eqnarray}
Assuming that the stream function is continuous across the injection region then
implies that an amount 
\begin{equation}
\Delta I=I_H-I_\infty=-\frac{(\Omega_H-\Omega_{+}-\Omega_{-})\psi}{2\pi}
\label{DI}
\end{equation}
of cross-field electric current must be supplied into the flux tube.
In situations where inertia is negligible, so that the inflow and outflow 
can be considered force-free, the current $\Delta I$ must be supplied by the
injection region. Under the assumption that the stream function is continuous 
a gauge can be found such that $A_t\simeq-\psi/2\pi$ to lowest order.  The 
potential drop along magnetic flux tubes in the injection region is then given 
by (see eq [\ref{EB}] and the text below),
\begin{equation}
\Delta V=-(\Omega_{+}-\Omega_{-})\psi/2\pi.
\label{DV}
\end{equation}
Once a solution $\psi(r,\theta)$ of the Grad-Shafranov equation is found, 
eqs. (\ref{DI}) and (\ref{DV}) can be solved to yield the angular velocity of
magnetic field lines on the horizon near the axis in terms of the magnetic 
flux, cross field current, and potential drop in the injection region:
\begin{equation}
\Omega_{+}={1\over2}\left[\Omega_H+\frac{2\pi(\Delta I-\Delta V)}{\psi}\right].
\end{equation}
In general, $\Delta I$ and $\Delta V$ will be determined by the conditions 
in the injection region.  Consequently, the angular velocity of magnetic lines on the
horizon and, hence, the efficiency at which energy is extracted from the hole 
are determined by the microphysics of the injection region.  The values of $\Delta V$
and $\Delta I$, the toroidal electric current in the interface separating the inflow and 
outflow, and the continuity 
of the stream function across this interface provide 4 boundary conditions for
the Grad-Shafranov equation governing the structure of the double transonic flow.
The values of the injected densities, $n_{inj\pm}$, enthalpies
, $h_{inj\pm}$, and poloidal velocities, $u_{p\pm}|_{inj}$, provide the six additional
boundary conditions required.  Thus, it seems that the entire structure is 
determined essentially by the physical conditions in the injection region and the 
regularity conditions at the critical surfaces alone.  No additional boundary conditions
are required on the horizon.  
This remains true in the limit of very small (albeit finite) inertia.  In particular,
when $\Delta I<<I_{GJ}$ and $\Delta V<<V_{max}$ , as expected in situations whereby the
plasma source is associated with a quasi-steady sparking gap, inside which the field aligned 
electric field $E_{||}$ is almost completely screened out, we obtain 
$\Omega_{+}\simeq\Omega_H/2$ near the axis, confirming earlier results (Blandford \& Znajek 1977, 
Phinney 1983).  Detailed analysis of a perturbed split 
monopole solution for a double transonic flow is presented in Beskin \& Kuznetsova (2000),
who reached similar conclusions.

\subsection{Inflow from a uniformly magnetized, massive torus}

\begin{figure}
\centerline{\epsfxsize=140mm\epsfbox{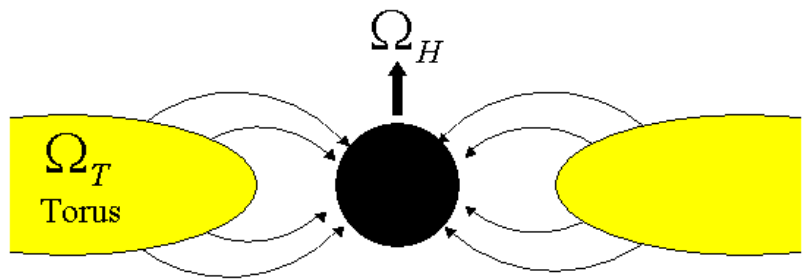}}
\caption{\small Inflow from a massive, infinitely conducting torus: In this example 
there is a direct link between the torus and the black hole.  The angular velocity 
of magnetic field lines that penetrate the horizon 
equals $\Omega_T$. Energy and angular momentum are transferred from the 
black hole to the torus when $\Omega_H>\Omega_T$, and in the opposite direction when
$\Omega_H<\Omega_T$.}
\end{figure}
As a second example we consider a situation in which the magnetic field lines extending
from the horizon are anchored to a massive torus.  This is shown schematically in figure 2.
We suppose that the mass flux along field lines connecting the torus and the 
black hole is small, and approximate the inflow as force-free.   This may be justified
in cases where angular momentum transfer from the hole to the torus results in a slightly
super-Keplerian rotation of the inner parts of the torus, thereby strongly 
suppressing accretion (van Putten \& Ostriker 2001).

For an infinitely conducting torus, the angular velocity of magnetic field
lines that originate from the torus equals the the angular velocity 
of the torus, viz., $\Omega_F=\Omega_T$. 
The torque exerted on the inner face of the torus by the black hole is 
obtained by integrating the angular momentum flux ${\cal L}^{r}$ given
by eq. (\ref{Lr}) 
over the section of the horizon which is threaded by magnetic field
lines that are anchored to the torus:
\begin{equation}
\tau=-4\pi\int_{\theta_H}^{\pi/2}{\sqrt{-g}{\cal L}^{r}d\theta}=
(\Omega_H-\Omega_{T})\int_{\theta_H}^{\pi/2}\frac{\Sigma}{\rho^2}
\sin\theta(F_{\phi\theta})^2 d\theta,
\label{torque}
\end{equation}
where eq. (\ref{ratio1}) has bee used.  Here $\theta_H$ is the angle of the 
last field line that connect the torus and the horizon.
In terms of the net poloidal magnetic flux associated with
the open field lines in the torus, $\psi_0$, we may write 
$\tau=(\Omega_H-\Omega_{T})f_H^2\psi_0^2/4\pi$.  Formally $f_{H}$ is defined through 
eq. (\ref{torque}).  It roughly represents the fraction of surface area of the horizon 
that is threaded by magnetic field lines emerging from the torus. 
It is seen that when $\Omega_H>\Omega_{T}$ angular momentum is transferred 
from the black hole to the torus, tending to spin up the inner face, whereas 
in the slowly rotating case ($\Omega_H<\Omega_{T}$) the black hole
receives angular momentum from the torus.  The associated power is 
given simply by $L=\Omega_T \tau$.

\subsection{Is there a causality problem?}
\label{sec:causality}

The question whether a force-free magnetosphere can exist and whether it is stable
has been the subject of a long debate (Punsly \& Coroniti 1990, 
Beskin \& Kuznetsova 2000;  Komissarov 2001,2003,2004b; Blandford 2001, 2002; 
Punsly 2003, 2004 van Putten \& Levinson 2003; Levinson 2004).  In 
particular it has been argued that a force-free black hole 
magnetosphere is not a causal structure and, therefore, physically excluded.
The claim made by Punsly and collaborators is that the fact that
the angular velocity, $\Omega_F$, of a force-free flux tube extending from the 
horizon is not a free parameter, but is determined by matching boundary conditions 
on the horizon and at infinity (as in the analysis of BZ and Phinney 1983) 
violates the principle of MHD causality.  The reason is that the inflowing magnetic wind must 
pass through the inner light cylinder before reaching the horizon and, therefore, 
cannot communicate with the plasma source region (see section \ref{sec:waves}).  Their 
main conclusion is that the use of the Znajek frozen-in condition on the horizon
to determine $\Omega_F$ is unphysical, and that $\Omega_F$ must be determined 
by the dissipative process that leads to ejection of plasma on magnetic field 
lines between the inner and outer Alfv\'en points.  This view seems to follow from 
the interpretation of the membrane paradigm, that the horizon can be viewed
as a conducting surface (see also Komissarov 2004b).
It has already been emphasized in the 
preceding section that the Grad-Shafranov equation is in general hyperbolic near the
horizon and requires no boundary conditions there, and that the Znajeck frozen-in 
condition is essentially a regularity condition on the fast critical surface.
Moreover, it has been argued that the injection region alone determines the
entire solution of the double transonic flow.
The question whether a force-free magnetosphere is stable and how it evolves
is nonetheless relevant.
Blandford (2001, 2002) suggested that any changes are communicated 
by means of a fast magnetosonic mode, that can propagate across magnetic
field lines at the speed of light and carry information about the toroidal magnetic 
field and poloidal current to the plasma source even beyond the light cylinder. 
Punsly (2003) argued that the fast mode can only carry displacement
currents, and therefore concluded that fast characteristics cannot transmit sufficient
information to affect the global structure.  Levinson (2004) in turn claimed that 
to second order the fast mode can affect the poloidal current (see section \ref{sec:waves} 
below for further details). 
The present view of this author, as explained below, is that although this 
is an interesting issue, it is of little relevancy to the causality problem.

To elucidate how a force-free magnetosphere responds to changes in spacetime, 
consider the following Gedunken experiment.
Suppose that a double transonic flow has been ejected along the rotation
axis of a Kerr black hole, and assume further that the flow is force-free everywhere 
except for the plasma source region where a small deviation 
from force-freeness must exist.   Imagine now that the angular momentum of the hole 
changes abruptly at time $t_0$ from $a_0$ to $a_0+\delta a$.  This will induce a 
change $-\delta \beta(r)$ in the angular velocity
of a ZAMO at any radius $r$.  The question then is how this variation in space-time
would affect the flow.  From eq. (\ref{j_GJ}) it is evident that a change 
\begin{equation}
\delta j^t(r)=-\frac{(\delta\Omega_F+\delta\beta)B_r}{2\pi\alpha^2}
\label{delta j}
\end{equation}
in the local electric charge density near the axis is required in order for the MHD flow to 
remain ideal, where $\delta \Omega_F$ is the resultant change in the angular velocity of the 
magnetic flux tube.  This is particularly true in the region between the inner Alfv\'en and
fast critical surfaces.  Whether the charge density adjusts by means of 
second order effects of the fast mode, as suggested by Levinson (2004; but cf. Punsly 2004), 
or otherwise is an interesting question by its own.  In our view it must adjust.  The 
crucial point is that any deviation of $j^t$ from the Goldriech-Julian value will 
induce a field-aligned electric field (i.e., ${\bf E}\cdot{\bf B}\neq0$), in violation
of the force-free assumption, that would tend to be screened out instantaneously
by the surrounding plasma.  Such a process involves most likely generation of longitudinal
modes that cannot be analyzed within the framework of force-free wave analysis anyhow.
Whether a quasi steady-state can be maintained in such a case is unclear to this author.
Recent time-dependent analysis of sparking gaps demonstrates that a response of this kind
might be oscillatory in nature (Levinson et al., in preparation).  The analysis outlined in 
section \ref{sec:frame-d} demonstrates that the magnetosphere remains force-free during the
course of evolution provided that sufficient plasma is present on field linse.

Now, if conditions in the plasma source are such that it always tends to be 
sustained in an approximately force-free state, in the sense that the 
cross field current $\Delta I$ and the potential drop across the plasma 
source region $\Delta V$ are negligibly small, then it follows from 
eqs. (\ref{I_H}) and (\ref{I_inf}) that as long as $\Omega_F\neq (\Omega_H+\delta \Omega_H)/2$,
the charge density in the gap will change with time sine $\vec{\nabla}\cdot \vec{j}\ne0$.
This will feedback on the potential drop $\Delta V$, and will lead to an evolution of 
the poloidal current and, hence, $\Omega_F$, until the value
$(\Omega_H+\delta \Omega_H)/2$ is exceeded\footnote{This assumes again
that the system can relax to a steady-state. Alternatively, it may 
undergoes large amplitude oscillations.}. 

\section{THE GLOBAL STRUCTURE}

The global structure of the magnetosphere determines the geometry of
magnetic flux surfaces and the location of regions where substantial departure from 
ideal MHD takes place.  The existence of such regions is essential in order to 
allow for (i) cross-field currents that are necessary for global current closure,
and (ii) dissipation of magnetic energy to account for the high entropy inferred 
in most astrophysical applications.  Moreover, the global structure determines through
which channels the BH rotational energy extracted is emitted.

In most astrophysical applications it is envisioned that the magnetic field threading the horizon
is supported by currents flowing in a disk or torus surrounding the black hole.  Two inherently
different disk magnetizations are discussed in the literature.  The first one consists of
what we term an open-field magnetosphere.   In the second one, termed closed-field geometry,
a large portion of magnetic field lines that thread
the horizon are anchored to the torus (Nitta et al. 1991; van Putten 1999, 2001; Uzdensky 2004b). 
This introduces a strong coupling between the black hole and the torus.  
In what follows we discuss those two magnetospheric configurations 
in some greater detail. 

\subsection{Open-Field Geometry}
\begin{figure}
\label{fig.open}
\centerline{\epsfxsize=140mm\epsfbox{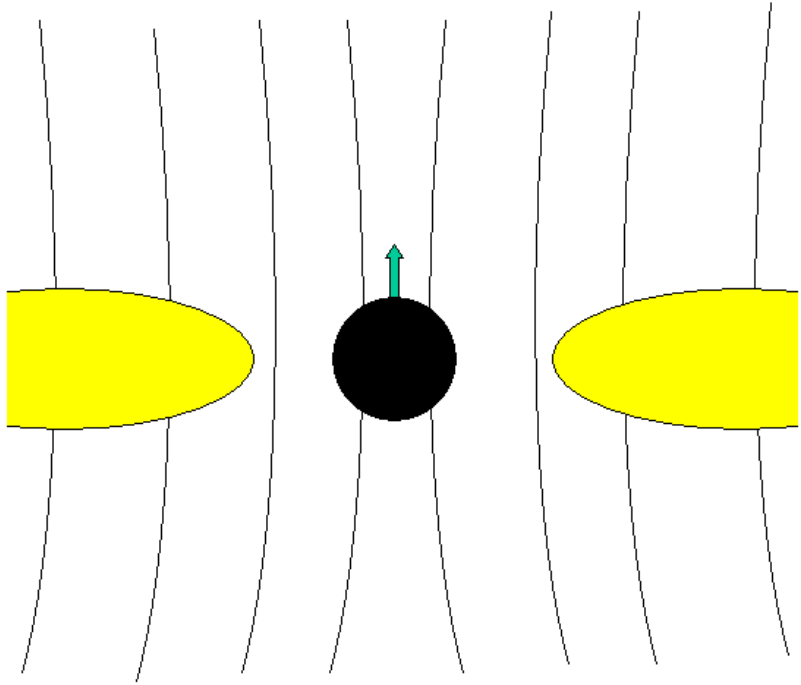}}
\caption{\small Schematic representation of an open-field magnetosphere}
\end{figure}
This configuration has been studied extensively (e.g., Blandford 1976; Lovelace 1976;
Phinney 1983; Macdonald 1984; Beskin \& Par\'ev 1993), and was the one employed originally 
by BZ in their model.  It is shown schematically in fig 3. 
In this configuration there is no direct link between the horizon and the 
disk.  In particular, all field lines that penetrate the horizon extend to infinity.
The energy extracted from the hole is then transported outward to the load 
along those field lines in the form of a Poynting flux.  The nature of the
load, where magnetic energy is presumably being converted to kinetic energy,
is not well understood at present.  The current closure path is also not well
understood.  It is conceivable that a current sheet may form on the equatorial
plan, thereby providing a return path.  This idea seems to be supported by recent 
numerical simulations (Komissarov 2004b).  If so, then a torque may be exerted on the 
inner parts of the disk by the cross-field currents.  This may give rise 
to enhanced dissipation, which has claimed to be inferred recently from 
X-ray observations of MCG-6-30-15 (Wilms, et al. 2001).  However, rapid heating
of the inner parts of the disk are expected also in the closed-field magnetosphere,
though for a different reason as explained below. 

\subsection{Closed-Field Geometry}
\label{sec:closed}
\begin{figure}
\label{fig.vacB}
\centerline{\epsfxsize=140mm\epsfbox{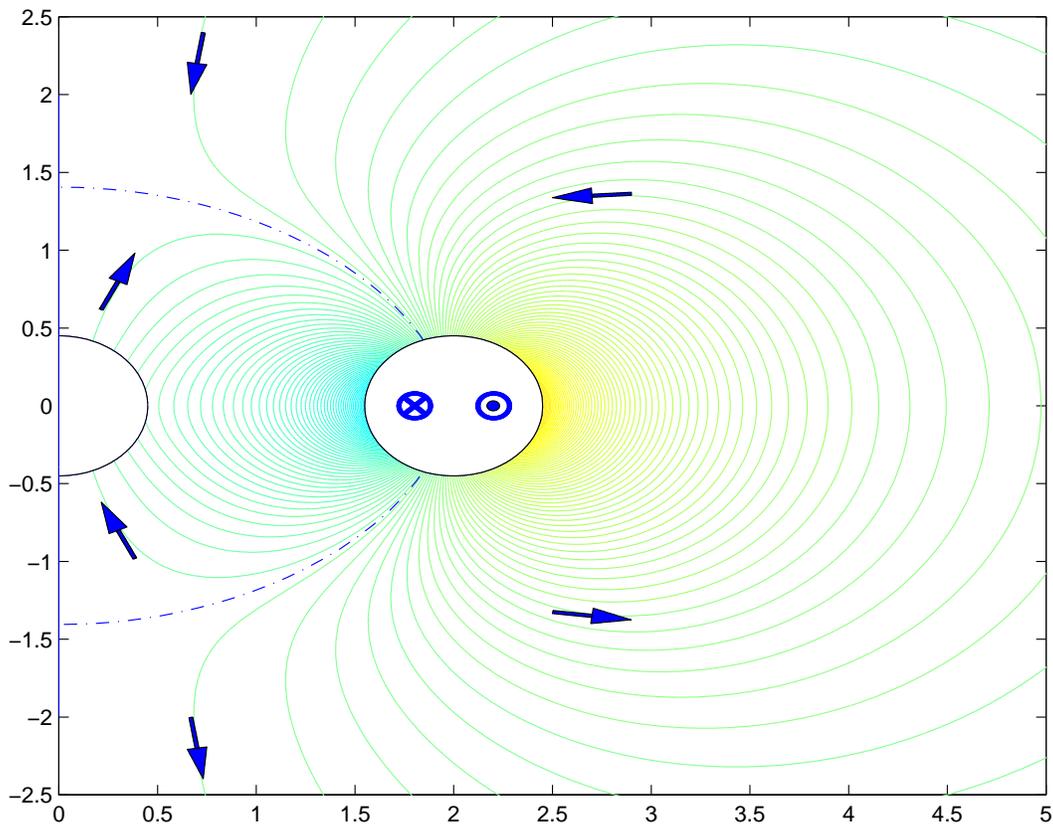}}
\caption{\small The poloidal topology of magnetic flux-surfaces in vacuum
produced by two counter-oriented current rings is shown.  The curent rings 
are located on the equatorial plan, one on the side of the torus
facing the black hole and the other one on the opposite side, as indicated.  
The dashed line is the separatrix between the flux-surfaces supported by the inner 
and the outer faces of the torus. Reprinted from van Putten (2004).}
\end{figure}
The basic properties of such a magnetosphere can be illustrated by matching a vacuum solution
to a uniformly magnetized torus.  Such a solution is topologically equivalent 
to a configuration produced by two counter-oriented current rings in the equatorial plane,
which in flat space-time can be calculated analytically.  
The resultant magnetic flux surfaces of this vacuo solution are exhibited in fig. 4.
A third current loop associated with the black hole equilibrium magnetic
moment in its lowest energy stat (van Putten 2001) has been added in the figure.
It is oriented antiparallel to the surrounding torus magnetosphere, 
facilitating an essentially uniform and maximal horizon flux at arbitrary spin-rates.
It only affects the solution very near the horizon.
As seen, there are two separated regions of closed field lines.   At large radii 
(compared with the radius of the outer current ring) the field quickly approaches 
a dipole solution.  In the inner region the field lines intersect the 
horizon, giving rise to a strong coupling between the black hole and the inner
face of the torus.  

\begin{figure}
\centerline{\epsfxsize=140mm\epsfbox{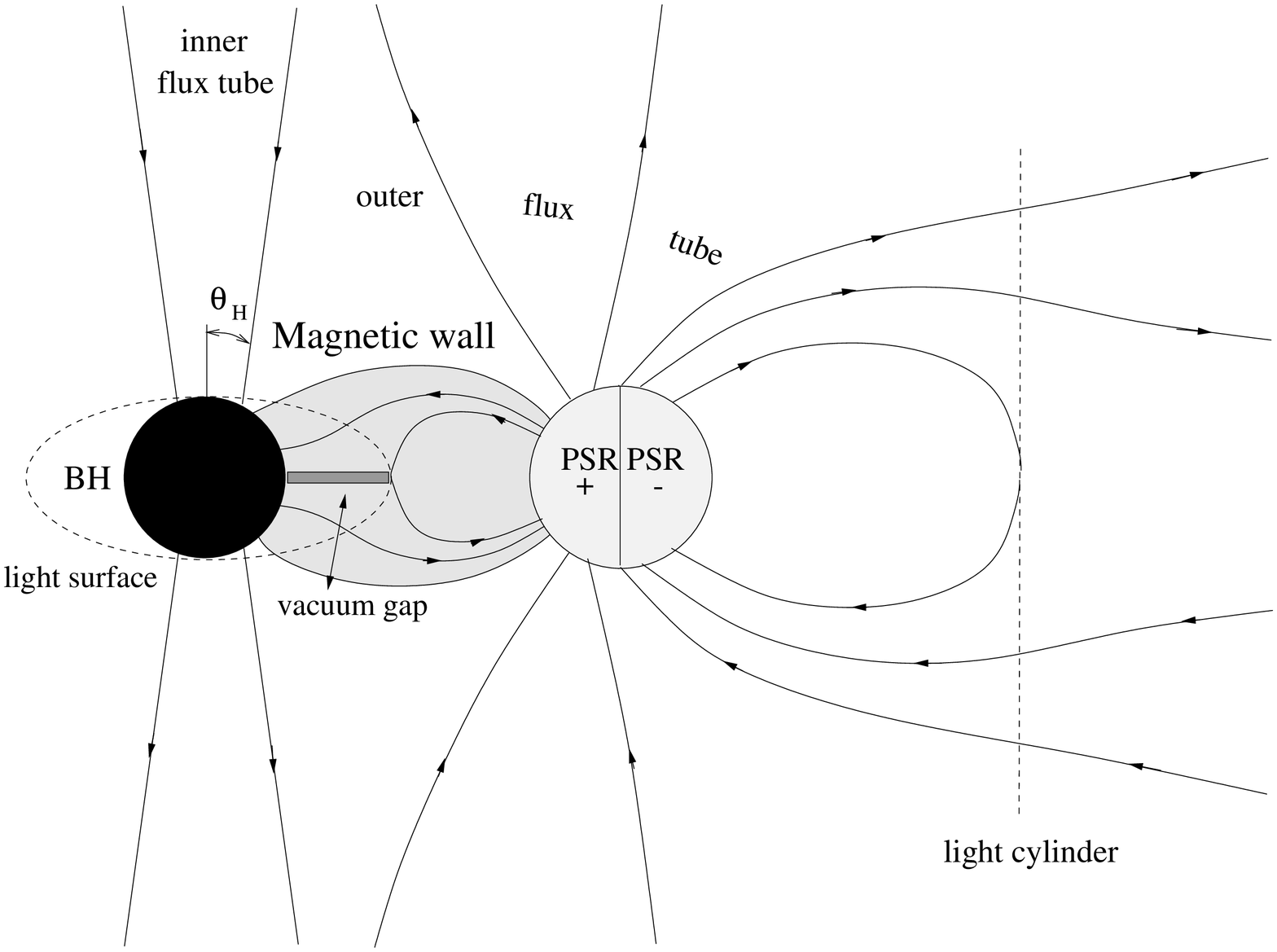}}
\vskip0.1in
\caption{\small Schematic illustration of the poloidal topology of the magnetosphere 
of a torus surrounding a rapidly rotating black hole. 
The direct link between the inner torus magnetosphere and the horizon results
in energy and angular momentum transfer from the black hole to the torus. 
This input is catalyzed by the torus into gravitational radiation, winds, thermal 
and MeV neutrino emissions.  The associated Maxwell stresses are mediated by poloidal 
currents, which close over a current sheet (marked as vacuum gap) in an annulus of vanishing 
magnetic field (as indicated).  A baryon-poor inner flux-tube serves as an artery for 
a {\em minor} fraction of black hole-spin energy. The inner and outer flux tubes are
separated by a charge and current sheet in the outflow section.  The dashed lines
indicate the inner and outer light cylinders.
The lifetime of the system is set by the lifetime of rapid spin of the black hole.
Reprinted from van Putten \& Levinson ApJ 584, 937 (2003).}
\label{fig.magneto}
\end{figure}

Now, the magnetic field inside the torus cannot be purely poloidal, because purely poloidal
fields are unstable and tend to decay completely in a few Alfven timescales 
(Markey \& Tayler, 1973; Flowers \& Ruderman, 1975; Eichler 1982).  However, by 
conservation of helicity, a twisted magnetic field does not
decay to zero, at least in the limit of ideal magnetohydrodynamics.
Instead, it will evolve into a new, 
stable configuration.  Recent 3D MHD simulations (Braithwaite \& Spruit, 2004) show that 
magnetic fields inside stars tend to develope a belt of twisted field lines that 
stabilize a dipolar field in the magnetosphere above the stellar surface.  This 
configuration appears to be stable over the resistive timescale, which is typically much 
longer than the canonical dynamical timescales (e.g., rotation
periods and accoustic timescales).  By topological equivalence 
in poloidal cross-section, it is therefore conceivable that the
magnetic field inside the torus is twisted as well, supporting
an overall torus magnetosphere which, from the outside, is 
consistent with a uniformly magnetized surface of the torus.

Like in pulsars (Goldriech \& Julian 1969; Michel 1982), the vacuum magnetosphere of the 
rotating torus is unstable.
By vacuum break-down, the flux-surfaces will evolve with electric charges to a
largely force-free state.  As a result, a magnetosphere develops which
consists of conductive flux-surfaces and magnetic winds.  This is shown 
schematically in fig 5.
In the limit of infinite torus conductivity, the flux-surfaces in the outer/inner torus 
magnetosphere assume rigid corotation with the outer/inner face of the torus.
A transonic inflow is expelled from the inner face along the open field lines to the
horizon.  For a rapidly rotating black hole we anticipate $\Omega_T<\Omega_H$, and so 
energy and angular momentum are transferred from the hole into the inner face of the torus
(see section 2.7), tending to spin it up.
Likewise, a transonic outflow is ejected from the outer face to infinity, resulting in 
a loss of energy and angular momentum of the outer face.  A quasi steady-state is quickly 
reached, whereby a flow of angular momentum mediated by shear forces between the
differentially rotating torus layers is established (van Putten \& Ostriker 2001).
The energy deposited in the torus is emitted predominantly in the form of gravitational
waves, owing to large deformations of the torus, as well as MeV neutrinos and baryon rich winds 
to infinity, resulting from the rapid heating of the torus by the friction between its layers
(van Putten \& Levinson 2003).   This is one of the major differences between the open-and-closed-
field geometries.

When the magnetic field strength inside the torus exceeds a few time $10^{15}$ G
it becomes dynamically important.  It is then anticipated that magnetic stresses that 
builds up inside the torus will lead 
to nonlinear, dynamical deformations of the torus.   The power spectrum of 
the induced mass moments is likely to be dominated by the several lowest multipoles. 
A naive estimate of the field strength above which the torus becomes
unstable to deformations may be obtained by equating the torque acting on a 
perturb current ring, owing to the mutual magnetic interaction between the 
two current rings, with the gravitational torque exerted by the central black hole
(van Putten \& Levinson 2003).  This yields a critical field strength of about 
10$^{16}$ G.  The presence of toroidal field components inside the
torus may somewhat alter this estimate.  Interestingly, this critical magnetic field 
corresponds to a spin down time of the order of tens of seconds for a stellar mass 
black hole, consistent with durations of long GRBs.  Full 3D GRMHD simulations are ultimately 
needed to study the dynamics of the torus (including the back-reaction of the black hole) 
in this regime.
Using a toy model, Bromberg et al. (2005) have recently calculated the evolution of the mass
moments of the deformed torus.  Their simulations reveal the existence of a nonlinear, 
oscillatory phase that sets in when the magnetic field approaches the critical
value.  The duration of this phase is found to be very long compared with the 
orbital period, for a certain range of parameters.  The deformations of the torus 
that result from this magnetic self-interaction should give rise to a burst of 
gravitational wave emission with durations of several to several tens of seconds (see Bromberg 
et al. 2005 for analysis of the gravitational wave spectrum).
For reasonable parameters we anticipate such sources to be detectable by 
LIGO and VIRGO out to a distance of 100 Mpc and even larger (van Putten et al. 2004)

The powerful wind driven by the pressure gradients in the surface layers of 
the hot torus passes through the outer Alfv\'en point. This results in the opening of
some magnetic field lines in the outer layers of the inner torus magnetosphere. 
Because the torus is rotating and magnetized, the ejection of the wind 
is partially anisotropic. Specifically, mass flux is generally
suppressed along magnetic field lines that are inclined toward the rotation axis,
and enhanced along field lines that are strongly inclined away from the axis
(Blandford \& Payne 1982; Romanova et al. 1997), owing to centrifugal forces.  
The details of the outflow depend on the heating and cooling rate of the corona,
and on its structure.   We speculate that
large pressure gradients in the corona would tend to push matter along
some of the magnetic field lines originally connected to the horizon,
and that a combination of buoyancy and centrifugal forces may subsequently give rise to a twist 
of these field lines, some of which may ultimately fold and open to form an open magnetic flux 
tube near the rotation axis that extends from the horizon to infinity.
The resulting structure consists of two coaxial flux tubes with opposite magnetic orientation,
as shown in fig 5.
The inner and outer flux tubes are separated by a cylindrical
current and charge sheet that accounts for the jump in the electric and 
magnetic fields across the interface.
The lower section of the inner/outer flux-tube which connects to the horizon has,
instead, a parallel orientation between the poloidal magnetic field.
In the perfect MHD limit, the properties of the interface are described by jump
conditions as follow from Maxwell's equations.  The surface charge density is given by
\begin{equation}
4\pi\sigma_e=[\Omega r\sin\theta B_r]= r\sin\theta\Omega_T B_{r+}
-r\sin\theta(\Omega_H/2) B_{r-},
\end{equation}
where $B_{r+}$ ($B_{r-}$) denotes the radial magnetic field near the interface 
in the outer (inner) flux tube, and the poloidal and 
toroidal surface currents by,
\begin{eqnarray}
\begin{array}{rl}
4\pi J^r&=[B_\phi]=\left[\frac{\Omega r\sin\theta B_r}{v^r}\right],\\
4\pi J^{\phi}&=-[B_r].
\end{array}
\label{EQN_IP}
\end{eqnarray}
The poloidal current (\ref{EQN_IP}) results beyond the
Alfv\'en point, where the wind transports angular
momentum outwards to infinity.
The outflow in the inner tube is expected to be relativistic ($v^r=1$).  
If the baryon rich outflow from the torus also becomes relativistic, then 
the latter equation implies that $J^r=\sigma_e c$, namely the poloidal current 
in the boundary layer is solely due to the outflowing surface charges. 
This current sheet is a potential site for reconnection of magnetic field lines,
which would convert magnetic energy in the inner flux tube into kinetic energy.  

\section{NON-STATIONARY EFFECTS}
The discussion in the preceding sections focused on static magnetospheres.
The question of whether these structures are globally stable and how they evolve 
can only be addressed using time dependent analysis.  This is the ultimate goal
of general relativistic MHD (GRMHD) simulations, that only very recently became 
feasible.  Simple analytic analysis
is nonetheless important, both to give us some insight into this complicated problem
and to provide test cases for the simulations whenever possible. 

There are two crucial issues that we think need to be clarified.
The first one is the role of linear waves in the response of a 
magnetized flux tube to local changes.   The second and most important one,
is the role frame dragging plays in the global evolution
of a force-free magnetosphere.  We discuss these issues in tern in 
what follows.

\subsection{Small amplitude waves}
\label{sec:waves}
The system of GRMHD equations admits in general 4 types of modes - slow, Alfv\'en, 
fast and entropy, that can most easily be identified from a linear 
perturbation analysis in a homogeneous background.  There is ample literature on the properties
of those waves (for recent accounts see e.g., Anile 1989; Uchida 1997a,b; 
Komissarov 1999, 2002; Punsly 2001; van Putten 2004).  
The linear modes also define the critical surfaces of a static transonic flow as discussed 
in section \ref{sec:GS} above. In essence, each such surface acts as a one-way membrane for waves of the 
corresponding type.  Whenever the flow is locally perturbed, only information associated
with subcritical modes can be transmitted backwards to the source.  
In particular, no information can be transmitted from the regions beyond the inner and
outer fast critical surfaces, meaning that the horizon of a Kerr black hole is not in 
causal contact with the injection region.   The question then arises, how information 
about the state of the black hole is transmitted through a magnetized flux tube.

In the force-free limit only two physical modes remain, the Alfv\'en and fast 
modes (e.g., Uchida 1997b; Komissarov 2002).  As explained in section \ref{sec:FF},
in this limit the Alfv\'en surface coincides with the light cylinder, whereas the 
fast critical surface approaches the horizon.  This fact has led Blandford (2002) to propose
that changes near the horizon can be communicated by means of the fast mode, a proposal
that has been questioned by Punsly (2001; 2003).  Below, we re-examine this problem.
The most detailed investigation of force-free waves in Kerr spacetime is
presented in Uchida (1997a,b).  This author derived a set of PDEs for the Lagrangian displacement
of linear perturbations.  Using a WKB approximation he then  
solved it to lowest order.  His results confirm that to lowest order the fast wave is
electromagnetic, in the sense that it propagates along null geodesics and is purely transverse.
The Alfv\'en mode has, in general, a longitudinal component.  It also propagates at the speed of light,
but only along poloidal magnetic field lines.   It is these properties of the force-free waves (see Punsly
2003) that are at the base of the causality dispute discussed in section 
\ref{sec:causality}.   As we shall now argue, the response of a flux tube to changes in spacetime
involves higher order effects, which are not accounted for in the lowest order geometric optical 
approximation.  To illustrate this, consider a static flux tube extending from the horizon 
of a Kerr black hole.  Using eq. (\ref{j_GJ}) the GJ 
charge density near the axis, as measured by a ZAMO, can be expressed 
in terms of the $\theta$ component of the ZAMO electric field as,
\begin{equation}
\rho_{GJ}=\frac{E_\theta}{2\pi\tilde{\omega}},
\label{rho_GJ}
\end{equation}
where $\tilde{\omega}$ is the cylindrical radius, as defined below 
eq. (\ref{metric}).  Suppose now that the system is transformed 
from this initial state to a slightly different state, e.g., 
due to a change in the angular momentum 
of the black hole by a small amount $\delta a<<a$.  The change in the angular
velocity of the flux tube, as measured by a ZAMO, will be accompanied by a 
change $\delta E_\theta$
in the electric field, and a corresponding change in the GJ charge
density:
\begin{equation}
\delta\rho_{GJ}=\frac{\delta E_\theta}{2\pi\tilde{\omega}}.
\label{del-rho}
\end{equation}
Consequently, the perturbed GJ charge density is associated with changes on scales
comparable to the dimension of the flux tube.  This means that the adjustment
of the magnetosphere to global changes cannot be analyzed within the framework
of the geometric optical approximation, as attempted by Punsly.  It is not even
clear whether for such long wavelength perturbations decomposition into 
MHD modes is possible.  To be more precise, 
the perturbed charge density associated with any short-wavelength disturbance 
of some static solution is given by $\delta\rho_e={i\bf k}\cdot\delta{\bf E}/4\pi$, 
where ${\bf k}$ is the corresponding wave vector.
Thus, any changes in the charge density required for adjustments of the angular 
velocity of an evolving magnetic flux tube would appear only to order 
$(k\tilde{\omega})^{-1}$, which is neglected in the lowest order geometric optical 
approximation.  The fact that the fast mode is
purely transverse to lowest order does not by itself imply that
a force-free magnetosphere cannot respond to changes beyond the inner light
cylinder.  It simply means that account of higher order terms in the linear perturbation
analysis is mandatory for exploring such effects.   
Such an attempt is presented in (Levinson 2004), who confirms that to lowest
order the fast mode is indeed electromagnetic.  However, he finds (but cf. Punsly 2004) 
that to second order the fast mode have a longitudinal component, and that the
perturbed electric charge density beyond the light cylinder approaches $\delta \rho_{GJ}$, 
as given in eq. (\ref{del-rho}).  

\subsection{The frame dragging dynamo}
\label{sec:frame-d}
To gain some insight into the role of frame dragging 
in the evolution of a force-free flux tube,
let us first express Faraday law in terms of the ZAMO fields $B_\phi$, $E_r$, and $E_\theta$,
and the potential $A_\phi$, which will be used as our free variables. In Boyer-Lindquist
coordinates eq (\ref{F=0}) gives,

\begin{equation}
B_{\phi,t}+\frac{\sqrt{\Delta}}{\rho^2}
(\rho\alpha E_\theta)_{,r}-\frac{1}{\rho^2}(\rho\alpha E_{r})_{,\theta}
=\frac{\sqrt{\Delta}}{\rho^2}(\beta_{,r} A_{\phi,\theta}-\beta_{,\theta} A_{\phi,r}).
\label{M5}
\end{equation}
The term on the R.H.S of the last equation, which is absent in flat spacetime, represents
the effect of frame dragging.  As seen it only couples to the poloidal magnetic field,
and can be interpreted as a driver or a dynamo term.  This term cannot be gauged away 
owing to the fact that frame dragging is differential.  Physically, this can be attributed to the 
fact that the angular velocity of a ZAMO depends on radius, and so magnetic surfaces appear
to be in differential rotation in the ZAMO frame, thereby giving rise to a potential drop
along magnetic surfaces that tends to be screened out when sufficient plasma is present.
Let us explore further the role of this driving term in the evolution of a flux tube. 

Consider the evolution of a force-free magnetosphere of a slowly rotating black hole.
To derive the equations for the evolving electromagnetic field, we linearize Maxwell equations 
(\ref{F=j}) and (\ref{F=0}), using the hole angular momentum $a$ as the small parameter.  
We suppose that initially the magnetosphere is non-rotating, and that the
magnetic field can be described by the vacuum Wald solution (Wald 1974).  
To second order in $a/M$ the initial solution reads:
\begin{equation}
A_\phi=(B/2)r^2\sin^2\theta,
\label{Aphi-0}
\end{equation}
$B_\phi=E_r=E_\theta=0$.  To this order the force-free condition (\ref{FF}) reduces to

\begin{eqnarray}
\sqrt{\Delta}j_r=-r\cot\theta j_\theta,\label{FFb}\\
\sqrt{\Delta} E_{\theta}=r\cot\theta E_r,
\label{FFa}
\end{eqnarray}
where $j_a$ denotes the components of the ZAMO poloidal current, 
which are related to the Boyer-Linduist current through:
$j_r=(\rho/\sqrt{\Delta})j^r$, $j_\theta=\rho j^\theta$.
The $r$ and $\theta$ components of eq. (\ref{F=j}) yield,
\begin{eqnarray}
- E_{r,t}+\frac{\alpha}{\rho\sin\theta}(\sin\theta B_{\phi})_{,\theta}
=4\pi\alpha j_r,\label{M1}\\
E_{\theta,t}+\frac{\alpha}{\rho}(B_{\phi})_{,r}=
-4\pi\alpha j_\theta,\label{M2}
\end{eqnarray}
From equations (\ref{M5}) - (\ref{M2}), we obtain a differential equation 
for the electric field $E_\theta$ (see Levinson 2004 for further details):
\begin{equation}
(1+\Delta\tan^2\theta/r^2)E_{\theta,tt}+\frac{\sqrt{\Delta}}{r^2}(\sqrt{\Delta}
B_{\phi,t})_{,r}-\frac{\Delta}{r^3\cos\theta}(\sin\theta B_{\phi,t})_{,\theta}=0,
\label{eq-psi}
\end{equation}
with
\begin{equation}
B_{\phi,t}=\frac{\sqrt{\Delta}}{r^2}\left[-(\sqrt{\Delta}E_\theta)_{,r}
+\frac{\sqrt{\Delta}}{r}(\tan\theta E_{\theta})_{,\theta}
-\frac{6MBa}{r^2}\sin\theta\cos\theta\right].
\label{B_Tt}
\end{equation}
The last term on the R.H.S of eq. (\ref{B_Tt}) accounts for the differential frame 
dragging of the initial magnetic field, that is, $\beta_{,r}A_{\phi,\theta}-\beta_{,\theta}A_{\phi,r}
=-(6MBa/r^2)\sin\theta\cos\theta$ to second order in $a/M$.
Now, since the magnetosphere is initially non-rotating, no electric current is flowing 
in the system, viz., $j_r(t=0)=j_\theta(t=0)=0$.
To examine how the poloidal current is generated, we take the time derivative of 
eq. (\ref{M1}), and employ equations (\ref{eq-psi}) and (\ref{B_Tt}) to obtain
near the rotation axis (that is, at small angles),
\begin{equation}
j_{r,t}(t=0)\simeq -12BMa{\sqrt{\Delta}\over r^5}\cos^2\theta.
\end{equation}
Consequently, the poloidal current is driven solely by the frame dragging dynamo; the
only assumption being made is that there is sufficient plasma in space to allow the
condition ${\bf E}\cdot{\bf B}=0$ to be satisfied everywhere.  Note that the poloidal 
current is generated initially everywhere in space and not only on the horizon,
in contrast to the case of a Faraday disk, as in Punsly's (2001) waveguide model 
(cf. Komissarov 2003).  This clearly 
shows that the interpretation of the horizon as a unipolar inductor is inappropriate; 
it is the gravitomagnetic effect that is responsible for the adjustment of the 
magnetosphere to changes. A similar conclusion was drawn earlier by Komissarov (2003, 2004b).  
Equation (\ref{eq-psi}) is rather complicated.  Approximate, analytic solutions can be obtained 
near the horizon (Levinson 2004)\footnote{There is a typo in eq. (24) of Levinson (2004).
It should read: $B_{T,\tau}=x2M^2\sin\theta\left(f_0-\frac{3Ba}{8M^2}\sin2\theta\right)$.}.  
For the case considered here we obtain in the region where $\alpha<<1$, and for times 
$t<2M\ln(2M^2/\Delta)$: \begin{eqnarray}
E_{\theta}=\frac{3Ba}{8M^2}\sqrt{\Delta}\sin2\theta[\cosh(t/2M)-1],\\
B_{\phi}=-\frac{3Ba}{8M^2}\sqrt{\Delta}\sin2\theta\sinh(t/2M),\\
j_r=-\frac{3Ba}{4M^3}\sqrt{\Delta}\cos^2\theta \sinh(t/2M).
\end{eqnarray}
As seen, the perturbed force-free field grows exponentially with an e-folding time $2M$.
Note that after time $t\sim 2M\ln(2M^2/\Delta)$, at which the above solution is no 
longer valid, the toroidal magnetic field and the current 
evolve close to their steady-state values, viz., $B_T\sim (\Omega_H/2)F_{\phi\theta}\sin\theta$, 
$I=B_T/2$.  We conjecture that after this time the system will reach a steady state. This
suggests that the magnetosphere inside the ergosphere evolves to a steady-state solution
over a few dynamical times.  

\subsection{Numerical Simulations}
There have been several attempts in recent years to perform numerical simulations
of the Blandford-Znajek process.  They all show the tendency of the magnetosphere
to evolve towards a stable steady state.  In the GRMHD simulations reported in Koide et al. (2002)
and Koide (2003), the initial magnetic field configuration is described
by the Wald solution.
The plasma around the black hole has initially zero momentum, a uniform mass density, and
Alfv\'en velocity of 0.983 c (Koide 2003).  This initial condition is similar to the one invoked 
in section \ref{sec:frame-d}, except that plasma inertia is not neglected and 
the black hole in the simulations is nearly maximally rotating.  The simulations clearly show
the generation of a toroidal magnetic field by the frame dragging dynamo inside the ergosphere,
and the consequent decrease of the specific energy of the plasma until it becomes negative.
The twist of the magnetic field lines seems to be propagating outward, carrying energy and angular momentum 
on account of the black hole rotational energy.  The simulations reported in Koide et al. (2002) 
run for a time of about several $r_s/c$
after which the code crashes.  This run is, unfortunately, not sufficiently long for the system to reach a 
steady-state.  Nonetheless, it does show that the magnetosphere inside the ergosphere evolves 
over a few dynamical times, consistent with the analytic result derived in section \ref{sec:frame-d}.
Komissarov (2001) performed 2D, time dependent numerical simulations of
the evolution of a force-free, axisymmetric monopole configuration around a Kerr black hole.
He found that the solution quickly settles to a stable steady state, with the angular velocity 
of magnetic field lines approaching $\Omega_H/2$ (meaning maximum extraction efficiency),
thereby confirming the results of BZ.  He later generalized the numerical model 
to include inertial terms (Komissarov 2004a), and demonstrated that the 
inertia of the plasma near the event horizon is dynamically unimportant, and that 
the force-free limit is a good approximation for magnetically dominated flows, at least for the monopole
case.  The simulations also show the development of a double transonic flow that extends beyond the 
outer fast critical surface.   In a later paper (Komissarov 2004b), this author explored more realistic
configurations.  He also included a prescription for electric resistivity that enabled him to
incorporate dissipative magnetospheric regions (current sheets) in the numerical model.   
He confirms his earlier conclusions, that a stable steady state is reached whereby energy 
is extracted electromagnetically.  He finds, however, that in certain magnetospheric 
configurations, current sheets may form on the equatorial plane.  Such regions may provide
a path for the return current of the global circuit.   Moreover, in certain cases collimation
of the outflow is not seen up to a few tens of gravitational radii.

In another set of numerical experiments (Hirose et al. 2004; McKinney \& 
Gammie, 2004; De Villiers et al. 2005),
the evolution of a weakly magnetized torus surrounding a Kerr black hole has been examined.
It is generally found that the magnetic field in the torus is initially amplified via
the MRI and becomes turbulent.  A funnel region near the rotation axis of the hole is 
identified in those simulations, in which the magnetic field appears to be ordered and 
nearly force-free.   This region is well described by the BZ model ( McKinney \& 
Gammie, 2004), and it appears that energy is being extracted along those force-free flux tubes.
However, the overall energy flux (integrated over the horizon) is dominated by the enthalpy
of accreted plasma.  This is not surprising in view of the initial and boundary conditions 
invoked.   Nonetheless, the presence of a magnetically dominated polar region along which 
a significant fraction of the accreted energy is channeled into a Poynting flux jet is 
interesting and of direct astrophysical relevance. 

The conclusion to be drawn from the numerical
experiments described above is that the magnetosphere of a Kerr black hole is a stable, causal
structure, and that its evolution is driven by the frame dragging dynamo.

\section{CONCLUSION}
\begin{itemize}
\item The system of static, axisymmetric ideal MHD equations in Kerr geometry is characterized 
by five quantities conserved along magnetic flux surfaces; the specific
energy and angular momentum, the angular velocity of magnetic lines, the ratio of particle
and magnetic fluxes, and the entropy.  The
stream function that defines those flux surfaces obeys a second order, nonlinear PDE
that involves the five invariants.  

\item The specific energy of a given magnetic surface that 
extends down to the horizon becomes negative when two conditions are satisfied: 
(i) its angular velocity is larger than zero 
and smaller than that of the black hole, and (ii) the corresponding Alfv\'en point is located 
inside the ergosphere.  Energy and angular momentum can be transmitted from the horizon
outwards along such a negative energy flux tube, on account of the black hole rotational
energy.  The rate at which energy is extracted from the hole depends on the strength of
the magnetic field near the inner fast critical surface, and the angular velocity of the 
flux tube.  For typical astrophysical parameters, the total power that can be 
extracted is sufficient to 
account for the luminosities exhibited by relativistic systems, e.g., AGNs, GRBs, 
microquasars, provided the efficiency is high.

\item Transonic solutions must pass through several singular surfaces.
Of most interest are the slow magnetosonic, Alfv\'en and fast magnetosonic surfaces, on which 
the bulk velocity of the flow equals the velocity of the corresponding linear mode.  There are two
sets of such critical surfaces, the inner one associated with a transonic inflow into the hole,
and an outer one associated with a transonic outflow to infinity.  Each surface acts as 
a one-way membrane to the corresponding mode, and determines what information can be 
communicated backward to the injection region.  The entire structure of the flow is determined
by appropriate boundary conditions and the regularity conditions
on the critical surfaces.  This is true also in the limit of zero inertia. 

\item Double transonic 
structures that extend from the horizon to beyond the outer fast critical surface, as in 
the applications to astrophysical jets, must contain a region where the ideal MHD condition
is violated.  This region serves as a plasma source and 
must be located somewhere between the inner and outer Alfv\'en 
surfaces.  The global double-transonic flow structure is controlled entirely by the 
micro-physical conditions in the plasma source and the regularity conditions.  
In the force-free case the deviation
from force-freeness in the plasma source may be very small, and in any case 
is required only to ensure the continuity of electric current along magnetic field lines.  
In some configurations, as in the closed-field geometry discussed in section \ref{sec:closed},
a cylindrical current sheet may separate the inner, double transonic flow and the outer
magnetosphere.

\item The extracted energy may be released through various channels, depending on the global
structure of the magnetosphere.  In open-field magnetospheres, where there is no direct link between
the black hole and the surrounding disk, the extracted energy is ultimately channeled along the rotation
axis in the form of a Poynting flux dominated outflow.  In closed-field magnetospheres, a
significant fraction of magnetic field lines that penetrate the horizon are anchored to the 
surrounding torus.   In such configurations the major fraction of the extracted energy is
transfered to the torus, and only about 0.1 \% are channeled along the rotation axis.
The energy deposited into the torus may result in strong gravitational wave emission and
baryon rich winds.  The ram pressure of the baryon rich wind that ensheath the inner jet 
can provide a means for collimating the inner jet.

\item The global evolution of a magnetosphere is governed by frame dragging.  It is 
the fact that the ZAMO angular velocity varies with radius that causes the appearance
of field-aligned electric field in vacuum (or a starved magnetosphere more generally).
In regions where sufficient plasma is present the parallel electric field tends to be 
screened out.  The adjustment of the electric charge to changes in spacetime is 
instantaneous essentially, and is dictated solely by the requirement that the 
condition ${\bf E}\cdot{\bf B}=0$ is preserved in the course of evolution.
The term $\hat{{\bf m}}{\bf B}\cdot{\bf \nabla}\beta$ that appears in the homogeneous
Maxwell equations, where ${\bf B}$ is the ZAMO poloidal
magnetic field, -$\beta$ is its angular velocity, 
and $\hat{{\bf m}}$ is a unit vector in the $\phi$ direction,
can be treated as a driving term that generates
the toroidal magnetic field and poloidal electric current.

\end{itemize} 

\indent{ACKNOWLEDGMENT}

I thank J. Bekenstein, V. Beskin, S. Komissarov, Y. Lyubarsky, J.C. McKinney, 
and M. van Putten for useful comments.  
This work was supported by an ISF grant for the Isreali Center for High Energy Astrophysics

\hfill
\break

\end{document}